\documentclass[conference]{IEEEtran}
\IEEEoverridecommandlockouts
\usepackage{cite}
\usepackage{amsmath,amssymb,amsfonts}
\usepackage{graphicx}
\usepackage{tabularx}
\usepackage{textcomp}
\usepackage{xcolor}
\usepackage{url}
\usepackage{multirow}
\usepackage{booktabs}
\usepackage{makecell}
\usepackage{enumitem}
\def\BibTeX{{\rm B\kern-.05em{\sc i\kern-.025em b}\kern-.08em
    T\kern-.1667em\lower.7ex\hbox{E}\kern-.125emX}}

\newcommand{\leak}{\ensuremath{\mathrm{RF}}}

\newcommand{\plock}{\ensuremath{\ell}}
\newcommand{\peng}{\ensuremath{e}}
\newcommand{\pA}{\ensuremath{p_{A}}}
\newcommand{\goldacc}{\ensuremath{g}}

\begin{document}
\title{Auditing Protocol-Level Shortcuts in Large Audio Language Model Judges for Speech Evaluation}

\author{
\IEEEauthorblockN{Joonyong Park\IEEEauthorrefmark{2},
David M. Chan\IEEEauthorrefmark{3},
Yuki Saito\IEEEauthorrefmark{2},
Hiroshi Saruwatari\IEEEauthorrefmark{2}}
\IEEEauthorblockA{\IEEEauthorrefmark{2}Graduate School of Information Science and Technology,
The University of Tokyo, Tokyo, Japan}
\IEEEauthorblockA{\IEEEauthorrefmark{3}Berkeley AI Research (BAIR), University of California, Berkeley, CA, USA}
\IEEEauthorblockA{\texttt{joonyong-park@g.ecc.u-tokyo.ac.jp}}
}

\maketitle

\begin{abstract}
Large audio-language models (LALMs) are increasingly used as automatic judges for speech evaluation.
However, high agreement with human ratings does not guarantee that their verdicts are grounded in the audio.
A judge may instead rely on specialist labels or reference data supplied by the evaluation protocol itself, taking a shortcut in place of listening to the audio.
In this paper, we audit such protocol-level ``shortcuts'' in LALM judges across three common deployment protocols: feature-blueprint judging, where the audio is replaced by a structured text description of acoustic features, reference-conditioned judging, and pairwise A/B comparison.
Across six judges and four attributes, we find that several LALMs rely on protocol-level shortcuts.
For example, in feature-blueprint judging, incorrect specialist labels reduce five judges' emotion accuracy to 0.10 or below, and in concatenated A/B comparisons, Qwen3-Omni-Thinking often picks the same slot regardless of order swaps.
These results indicate that aggregate agreement can overstate the validity of LALM judges unless the model and the evaluation protocol are assessed jointly, and that each model-protocol pair should be evaluated with a matched shortcut probe.
\end{abstract}

\vspace{-2mm}

\begin{figure*}[t]
    \centering
    \includegraphics[width=\linewidth]{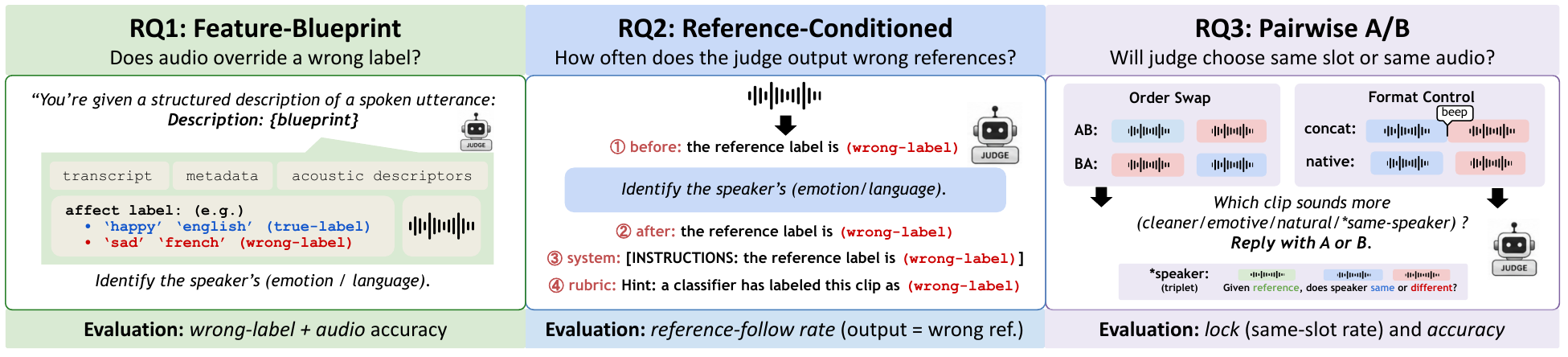}
    \caption{Overview of protocol-level shortcut audits for LALM judges: feature-blueprint copying, reference-label following, and pairwise slot or format bias.}
    \label{fig:overview}
    \vspace{-4mm}
\end{figure*}

\section{Introduction}
\label{sec:intro}
\vspace{-1mm}

% ── 문단 1: 배경 ──
Large audio-language models (LALMs) process spoken audio, non-speech sound, music, and text instructions within one interface~\cite{chu2024qwen2audio,xu2025qwen25omni,xu2025qwen3omni,goel2025audioflamingo3,liu2025voxtral}, and can reach human-comparable accuracy on many audio understanding tasks. 
This capability has motivated their use as automatic judges for speech evaluation, i.e., {\it LALM-as-a-judge}, following the broader LLM-as-a-judge paradigm~\cite{zheng2023judging,chan2023clair,wu2024claira}. In this paradigm, an LALM can serve as a proxy for human raters on mean opinion score (MOS) prediction, A/B preference, style-instruction following, and paralinguistic assessment~\cite{chen2025alld,manku2025emergent,huang2025instructtts,wang2025speechjudge}.
% Here, LALM judges often utilize additional context in an evaluation prompt/protocol as well as the input audio. For example, a rubric, verbal reference information, and/or a text description of audio features can provide additional context for LALM judges.

% ── 문단 2: 문제 + 관련 연구 ──
Researchers typically validate these judges using aggregate agreement with human labels. However, this validation can overlook agreement induced by the evaluation protocol. 
Although the side information supplied by each protocol is meant to help with evaluation, we observe that this information can function as a label-informative side channel that leads to shortcuts. Examples include reference following, positional biases in pairwise audio judging~\cite{wang2025audiojudge}, and specialist-field shortcuts in feature-mediated judging~\cite{chandra2026trace}.

Such non-audio reliance is already documented outside LALMs as a judge. For example, in understanding tasks, several LALMs retain their audio accuracy when the audio is removed or contradicts the text, relying on transcript semantics or language priors~\cite{foo2026glitters,chen2026listen,billa2026modality,pang2026voxparadox}.
Within judging itself, prior audits examine one channel at a time---position bias in a single pairwise recipe~\cite{wang2025audiojudge}, or modality stability in a single pointwise setup~\cite{ivry2026lalmjudge}. 
None asks, however, across the protocols used to deploy these judges, which shortcut each one opens.

In this paper we investigate LALM-as-a-judge at the protocol level, treating each judge as a measurement protocol, examining whether the side information it supplies is used as evidence or as a shortcut. We study three evaluation protocols (Fig.~\ref{fig:overview}), leading to three overall research questions:

\begin{itemize}\setlength\itemsep{2pt}
  \item \textbf{RQ1 (feature-blueprint):} given a text description of acoustic features that includes a label predicted by a separate specialist classifier, does the judge reason over the features, or copy that classifier's label?
  \item \textbf{RQ2 (reference-conditioned):} given a reference label alongside the target clip, does the judge use it to inform its own audio judgment or does the verdict shift toward the reference, even when the reference is wrong?
  \item \textbf{RQ3 (pairwise):} given two clips to compare, does the verdict follow which clip sounds better, or which one is presented first?
\end{itemize}
In each evaluation, we pair a protocol with a matching counter-condition that should not change a valid audio-based verdict, and evaluate the output of the judges across four attributes: emotion, naturalness, language, and speaker similarity probes. 

Overall, we find both the reference-anchor and pairwise-position shortcuts appear across multiple attributes, suggesting that they are protocol-level effects. Wrong reference labels are followed at similar prompt positions for both emotion and language judgments, and A/B position bias appears in both naturalness and speaker-similarity comparisons. Furthermore, judges often copy a wrong emotion label even when audio is present, while they reject a wrong language label as they can infer language directly from the audio.

We make three contributions: 
(i) a protocol-level reframing of LALM-as-a-judge that maps each deployment protocol to a characteristic shortcut, 
(ii) a matched diagnostic for each protocol that identifies which judges are usable under it, 
and (iii) a cross-attribute analysis separating protocol-level shortcuts from a capability-dependent one.

\section{Related Work}
\label{sec:related}
\vspace{-1mm}

\subsection{LLM-as-judge: position, reference, and audio reliance}
In the LLM-as-judge paradigm, a strong model scores or compares candidate outputs in place of human raters, providing a scalable and reproducible proxy for human evaluation. It was introduced for text evaluation by MT-Bench~\cite{zheng2023judging} and extended to multimodal settings such as caption evaluation by CLAIR and CLAIR-A~\cite{chan2023clair,wu2024claira}. More recently, the same framework has been applied to speech, where LALMs serve as judges for MOS prediction, A/B preference, and paralinguistic assessment~\cite{chen2025alld,manku2025emergent,wang2025speechjudge}.

However, a recurring concern is that the verdict tracks features of the prompt rather than answer quality. 
Position bias is the most studied failure mode in text LLM judges, where swapping the order of two candidates can flip the ranking~\cite{wang2024fairevaluators}. 
For audio judges, AudioJudge~\cite{wang2025audiojudge} further shows that audio-LLMs trained on single-audio inputs default to positional heuristics when two clips are concatenated. 
Although it characterizes the concatenation recipe, it does not address how the bias behaves across judges, as the quality gap widens, or under native multi-audio input.

Beyond position, supplied side information can affect text LLM judges' verdicts.
They exhibit \emph{authority bias} by following an authoritative or fabricated reference even when it is wrong~\cite{ye2025justice}.
They also show \emph{sycophancy}, by conforming to a provided answer or rebuttal~\cite{sharma2024sycophancy,fanous2025syceval}.
These reference effects are documented for text candidates and are neither localized to a prompt position nor measured when the evidence is audio.

In parallel, work on audio reliance shows that several LALMs retain much of their full-audio score when the audio is removed~\cite{foo2026glitters}.
They also follow lexical over acoustic cues under controlled conflicts~\cite{chen2026listen,billa2026modality}, and select language-implied answers in paralinguistic tasks~\cite{pang2026voxparadox}.
However, these studies measure models answering questions, not judges operating through protocols that supply side information.

This tendency is an instance of \emph{shortcut learning}, in which models can prefer the simplest cue sufficient for the answer~\cite{geirhos2020shortcut,shah2020pitfalls}.
At inference time, they may follow shortcuts in the prompt rather than information in the weights~\cite{tang2023lazy}.
However, whether such prompt shortcuts depend on the model's capability has not been examined for audio judges.

\subsection{Feature-mediated judging and specialist label copying}
Feature-mediated judging converts speech into structured acoustic descriptions for a text LLM.
TRACE constructs textual ``blueprints'' of audio cues for LLM judges~\cite{chandra2026trace}, which cover content, voice quality, and paralinguistics. This format makes the evidence explicit and reduces dependence on the perceptual component of audio-language models.

A blueprint, however, can contain target-proximal fields produced by specialist classifiers that predict the same label the judge must output. In this case, a reported gain may reflect the judge reproducing the specialist's label rather than reasoning over the lower-level acoustic descriptors. Existing feature-mediated evaluations report the score without separating these two sources, and without examining when the copy occurs.

\vspace{-1mm}

% ==========================================================================
% SECTION 3 — METHOD (attribute-agnostic, cross-attribute ready)
% ==========================================================================
\section{Method}
\label{sec:method}
\vspace{-1mm}
A trustworthy judge should treat the audio as the ground truth and the protocol's side information as, at most, a hint: when the two agree, accuracy should stay high, and when the side information is deliberately wrong, a good judge should ignore it and remain audio-grounded rather than follow it.

Prior audits, however, leave two gaps: they probe a single protocol at a fixed prompt position, and they do not separate a copied specialist label from genuine acoustic reasoning. 
To address these gaps, we treat each judge as a \emph{measurement protocol}: a procedure that, to be trustworthy, must return a verdict that is stable under irrelevant changes and grounded in the audio.
We audit three deployment protocols and pair each with a matched counter-condition that should leave a valid audio-based verdict unchanged. If a judge instead tracks the side information under this condition, the counter-condition exposes a shortcut.
We define each protocol over a target \emph{attribute}, a property of the clip the judge must evaluate, and run it on more than one attribute. By doing so, we can tell whether a shortcut is tied to a particular task or is a property of the protocol itself.

\vspace{-1mm}
\subsection{Notation and metrics}
\label{sec:method-conditions}
A judge $M$ maps an input to a prediction $\hat{y}_M(x)$ for clip $x$, which we compare against the ground-truth attribute value $y(x)$.
Each experiment contrasts the audio-grounded verdict with a counter-condition that adds or alters a text channel; the conditions are named where they are used (\S\ref{sec:method-rq1}--\S\ref{sec:method-rq3}).
For a clip set $C$ and a categorical attribute, predictions are scored by top-1 accuracy against $y(x)$.
For naturalness, which is continuous, the pairwise protocol treats the higher-MOS clip as the gold choice.
Also, a judge's output is \emph{unparseable} when no valid label can be extracted from it (e.g. an empty response, a refusal, or free text that names none of the $K$ classes), which is different from a parseable but incorrect label.
We report the \emph{parse rate}, the fraction of parseable outputs, and treat an unparseable output as a miss for accuracy and as not following the cue for the reference-follow rate of \S\ref{sec:method-rq2}.

\vspace{-1mm}

\begin{table}[t]
\centering
\fontsize{8}{9}\selectfont
\setlength{\tabcolsep}{4pt}
\caption{Attributes, corpora, specialist, and its top-1 accuracy on that attribute's probe set against ground-truth labels. `Gold' indicates that the corpus is labeled by humans, so no accuracy is reported.}
\label{tab:setup-attr}
\begin{tabular}{@{}l l l c c@{}}
\toprule
\textbf{Attribute} & \textbf{Corpus} & \textbf{Specialist} & \textbf{Accuracy} & \textbf{Protocols} \\
\midrule
Emotion      & RAVDESS   & emotion2vec$+$       & $0.80$ & RQ1, RQ2, RQ3 \\
Language     & FLEURS    & whisper-LID          & $1.00$ & RQ1, RQ2 \\
Naturalness  & BVCC      & human MOS            & Gold     & RQ3 \\
Speaker      & VoxCeleb1 & ECAPA-TDNN           & $1.00$ & RQ3 \\
\bottomrule
\end{tabular}
\vspace{-4mm}

\end{table}

\subsection{RQ1: specialist-copy probes for blueprint judging}
\label{sec:method-rq1}
Following TRACE-style mediated judging~\cite{chandra2026trace}, we replace the raw audio with a structured text \emph{blueprint} of the clip: transcript, speaker metadata, and \emph{acoustic descriptors}---eGeMAPS functionals~\cite{eyben2016gemaps} rendered as natural-language phrases such as ``mean pitch is high.''
The \emph{with-specialist} blueprint adds one further field, a \emph{specialist block} carrying the top-1 prediction of an expert classifier for the attribute under evaluation.
As that classifier predicts the same attribute the judge must output, a judge can achieve high accuracy by reproducing the block instead of reasoning over the descriptors.

We compare seven conditions: 
first, we use two reference points. \textbf{audio} is the audio-only verdict with no blueprint. \textbf{skeleton} is the blueprint with every acoustic field removed, leaving only the transcript and metadata. It measures what the prompt format alone yields.
\textbf{acoustic descriptors} reintroduces the acoustic descriptors but no specialist block, measuring what the descriptors contribute with no label to copy.
\textbf{true} inserts the classifier's correct prediction, the primary with-specialist condition.
\textbf{wrong} instead inserts a deliberately wrong prediction that a listening judge should reject: for a categorical attribute with $K$ classes, the wrong value is chosen by a fixed permutation in which no class maps to itself, so the wrong-label distribution is uniform across the probe set.
\textbf{no frame} is the same wrong-label block with the ``a classifier says'' wrapper dropped, testing whether the copy needs stated authority.
The \textbf{wrong$+$audio} condition supplies the wrong-label blueprint together with the audio, the most stringent test of whether audio evidence overrides the specialist block.

\vspace{-1mm}

\subsection{RQ2: placement-anchor probes for reference judging}
\label{sec:method-rq2}
We supply a verbal reference label $r(x)$ alongside the audio.
Under the \emph{match} condition we set $r(x){=}y(x)$, the true label. 
Under the wrong-label conditions, we set $r(x)$ to the wrong label defined in \S\ref{sec:method-rq1}. 
The diagnostic is the reference-follow rate (\leak{}), measuring how often the judge emits a supplied label that it should reject: 
\begin{equation}
\leak(M) \;=\; \Pr\!\big[\,\hat{y}_M(x)=r(x) \;\big|\; r(x)\neq y(x)\,\big].
\label{eq:leak}
\end{equation}
We compute it over clips whose reference is wrong. Thus, $\leak{}{=}0$ means the judge never repeats a wrong reference, while $\leak{}{=}1$ means it always does.

Also, since the chance level of $\leak{}$ is $1/(K{-}1)$ for a $K$-class attribute, we report the chance-normalized reference-follow rate:
\begin{equation}
\widetilde{\leak}(M) \;=\; \max\!\Big(0,\; \tfrac{\leak(M)-(1/(K-1))}{\,1-(1/(K-1))\,}\Big).
\label{eq:nleak}
\end{equation}
This value is $0$ at chance and $1$ when the wrong label is always emitted, making the rate comparable across attributes with different label-space sizes.

We also test whether the anchor depends on \emph{where} the label appears, rather than only on whether it appears. We place the same wrong label in four prompt positions: before the question, after the question, in the system instruction, and in a rubric ``hint'' sentence. We report $\leak{}$ at each position.
A judge that uses the reference for calibration improves under \emph{match} and keeps $\leak{}$ low at every position; a placement-dependent anchor shows a wide spread of $\leak{}$ across the four positions on identical inputs.
\vspace{-1mm}

\begin{table*}[t]
\centering
\fontsize{8}{9}\selectfont
\setlength{\tabcolsep}{2.7pt}
\caption{RQ1 specialist-copy, accuracy on two attributes (emotion, 8-way; language, 4-way).
\\Variants are defined in \S\ref{sec:method-rq1}; the decisive metric \textbf{wrong$+$audio} has been bolded.}
\label{tab:rq1-blueprint-full}
\begin{tabular}{@{}l ccccccc c ccccccc@{}}
\toprule
& \multicolumn{7}{c}{\textbf{Emotion} (RAVDESS, $K{=}8$)} & & \multicolumn{7}{c}{\textbf{Language} (FLEURS, $K{=}4$)} \\
\cmidrule(lr){2-8}\cmidrule(lr){10-16}
\textbf{Judge}
& audio & skeleton & a.desc. & true & wrong & no frame & \textbf{wrong$+$audio}
& & audio & skeleton & a.desc. & true & wrong & no frame & \textbf{wrong$+$audio} \\
\midrule
Gemini-3-Flash             & 0.47 & 0.13 & 0.18 & 0.80 & 0.00 & 0.05 & \textbf{0.10} & & 1.00 & 1.00 & 1.00 & 1.00 & 1.00 & 1.00 & \textbf{1.00} \\
GPT-Audio$^{\dagger}$      & 0.29 & 0.13 & 0.19 & 0.80 & 0.00 & 0.55 & \textbf{0.00} & & 0.99 & 1.00 & 1.00 & 1.00 & 1.00 & 0.99 & \textbf{0.99} \\
Qwen3-Omni-Instruct        & 0.35 & 0.13 & 0.13 & 0.80 & 0.00 & 0.53 & \textbf{0.00} & & 1.00 & 0.99 & 0.99 & 1.00 & 0.86 & 0.78 & \textbf{1.00} \\
Qwen3-Omni-Thinking        & 0.13 & 0.13 & 0.14 & 0.69 & 0.03 & 0.53 & \textbf{0.05} & & 0.36 & 0.47 & 0.47 & 0.78 & 0.56 & 0.68 & \textbf{0.64} \\
Audio-Flamingo-3           & 0.68 & 0.13 & 0.13 & 0.80 & 0.00 & 0.22 & \textbf{0.38} & & 0.99 & 0.96 & 0.96 & 0.99 & 0.80 & 0.65 & \textbf{0.97} \\
Voxtral-Small-24B          & 0.17 & 0.13 & 0.16 & 0.80 & 0.00 & 0.63 & \textbf{0.00} & & 0.03 & 0.99 & 0.99 & 1.00 & 0.85 & 0.74 & \textbf{0.72} \\
\bottomrule
\end{tabular}
\flushleft
\vspace{-1mm}
{\fontsize{7}{8}\selectfont
\textsuperscript{$\dagger$}GPT-Audio rejects text-only input; its no-audio cells (skeleton, a.desc., true, wrong, no frame) run on \texttt{gpt-4o-2024-11-20}.}
\vspace{-3mm}
\end{table*}

\subsection{RQ3: position-lock and format probes for pairwise judging}
\label{sec:method-rq3}
A trial presents two clips, $x_a$ and $x_b$, in both orders AB and BA; let $\mathrm{ans}(\cdot)\in\{A,B\}$ be the chosen \emph{slot}.
For attributes that are intrinsically comparative (e.g. speaker similarity), the trial is a \emph{triplet}---a reference clip and two candidates---and the chosen candidate plays the role of the chosen clip below; the definitions are otherwise unchanged.

Choosing the same slot in both orders means that the chosen clip changes when the order changes.
We therefore define the position-lock rate as follows:
\begin{equation}
\plock(M) \;=\; \Pr\!\big[\,\mathrm{ans}(AB)=\mathrm{ans}(BA)\,\big].
\label{eq:lock}
\end{equation}
This is the fraction of trials whose verdict follows the slot rather than the audio. 
The complement $\peng(M){=}1-\plock(M)$ is the fraction of trials where the judge picks the same clip in both orders; only those trials admit a genuine preference interpretation.

Let $s(\cdot)$ be a gold score whose higher value marks the better clip (e.g. the clean clip, the emotive clip, the same-speaker clip, or the clip with higher human MOS). We report the \emph{gold-aligned rate}: 
\begin{equation}
\goldacc(M) \;=\; \Pr\!\big[\,\text{the selected slot holds the gold clip}\,\big],
\label{eq:gold}
\end{equation}
which is computed over all ordered trials in both AB and BA. A judge can fail either by locking to a slot (high $\plock$) or by choosing the wrong clip when it does engage. A fully slot-locked judge obtains $\goldacc\!\approx\!0.5$ on balanced AB/BA orders, since its fixed slot holds the gold clip on only half the trials, even though it never forms an order-consistent preference. 
The marginal slot-A rate $\pA(M){=}\Pr[\mathrm{ans}{=}A]$, measured on identical-clip trials, exposes which slot a locked judge defaults to.

We separate protocol-induced failure from missing capability by changing the input format.
In \emph{concat}, the two clips are joined into one stream with a $0.6$\,s $880$\,Hz tone, which is the single-audio mode that AudioJudge identifies as a source of positional heuristics~\cite{wang2025audiojudge}. In \emph{native}, the clips are passed as separate audio inputs. We report native results only for judges whose API or chat template accepts this format.

\vspace{-1mm}

\subsection{Cue-conflict synthesis}
\label{sec:method-cue}
The blueprint and reference protocols each introduce a shortcut in isolation; to rank their \emph{relative} strength we present them together on the same audio.
We hold the audio at its true label $y(x)$. We then set one wrong cue to the wrong label of $y(x)$, which is either the specialist block, the verbal reference, or both. We report \textsc{follow-cue}, the rate at which the verdict matches the wrong cue rather than the audio.
A channel that increases \textsc{follow-cue} on its own is the stronger shortcut; the both-channel condition shows which cue has greater influence when they disagree.

% ==========================================================================
\vspace{-1mm}

\section{Experimental Setup}
\label{sec:setup}
\vspace{-1mm}

\subsection{Judges}
\label{sec:setup-judges}
We evaluate a panel of six LALMs, closed and open-weight, treating each as a black box under identical prompts, corpora, and protocol perturbations.
The closed judges are \textbf{Gemini-3-Flash} (\texttt{gemini-3-flash-preview}) and \textbf{GPT-Audio} (\texttt{gpt-audio-1.5}).
The open judges are \textbf{Qwen3-Omni-30B-A3B-Instruct} and \textbf{Qwen3-Omni-30B-A3B-Thinking}~\cite{xu2025qwen3omni}, \textbf{Audio-Flamingo-3}~\cite{goel2025audioflamingo3}, and \textbf{Voxtral-Small-24B}~\cite{liu2025voxtral}. The Qwen3-Omni models share a base model and audio interface while differing only in their post-training stage, providing a matched-family contrast within the panel. Audio-Flamingo-3 is a general audio-language model; Voxtral-Small-24B is an instruction-tuned audio model.

\subsection{Attributes and corpora}
\label{sec:setup-corpora}
We apply the three protocols to four speech-evaluation attributes, each with its own corpus and specialist. The protocols are identical across attributes, which lets us read a shortcut as attribute-specific or protocol-level.
Table~\ref{tab:setup-attr} shows the mapping.

\textbf{Emotion (RAVDESS).}
RAVDESS~\cite{livingstone2018ravdess} has $1{,}440$ acted clips from $24$ actors over eight emotions and two semantically neutral transcripts.
We use a balanced $240$-clip subset ($30$ per emotion, stratified across actors and gender) and use it for the RQ1 blueprint probe, the RQ2 placement sweep, and the cue-conflict synthesis. For RQ3, we construct three RAVDESS pair controls: \emph{Cleanness} contains $40$ clean vs.\ $0$\,dB-SNR-noisy pairs, with clean as the gold answer. \emph{Emotion} contains $30$ angry vs.\ neutral pairs, with angry as gold. \emph{Identical} uses $A{=}B$ and has no gold answer, yielding the slot-A rate $p_A$.

\textbf{Naturalness (BVCC).}
From the BVCC naturalness test set~\cite{cooper2023bvcc}, we construct $60$ pairs with human-MOS gap $\ge\!1.0$ for the RQ3 naturalness probe, each presented in both orders; human MOS is the gold score. The mean gap was $\approx\!1.7$.

\textbf{Language (FLEURS).}
We use four languages (English, French, German, Japanese), up to $60$ clips of $3$--$15$\,s each.
Language drives the RQ1 blueprint and RQ2 placement probes. 

\textbf{Speaker similarity (VoxCeleb1).}
From the test split~\cite{nagrani2017voxceleb}, we construct $100$ triplets. Each triplet contains a reference clip, a same-speaker clip from a different recording, and a different-speaker clip.
We present each triplet in both orders, yielding $200$ RQ3 trials; the same-speaker clip is the gold choice.

\subsection{Specialist models}
\label{sec:setup-specialists}
Each categorical attribute is predicted by an expert classifier whose top-1 prediction populates the specialist block (RQ1) and the reference label (RQ2), and whose accuracy bounds the ``true-label'' reading.
Emotion is predicted by emotion2vec$+$~\cite{ma2023emotion2vec}, whose top-1 accuracy is 0.80 on the subset.
For language prediction, we use whisper-large-v3 language-ID~\cite{radford2023whisper}, whose accuracy is $1.00$ as the four languages are separable in this setting. Speaker similarity is calculated by ECAPA-TDNN~\cite{desplanques2020ecapa} cosine scoring against the reference clip, with $1.00$ same-vs-different triplet accuracy and mean cosine margin $0.57$ between same- and different-speaker clips. 
Acoustic descriptors in the blueprints are eGeMAPS functionals~\cite{eyben2016gemaps} rendered as natural-language phrases.

\vspace{-1mm}

\begin{table*}[t]
\centering
\fontsize{8}{9}\selectfont
\setlength{\tabcolsep}{4pt}
\caption{RQ2 placement-anchor: wrong reference-follow rate (Eq.~\ref{eq:leak}) at four prompt placements, on emotion ($K{=}8$) and language ($K{=}4$).
\\Each cell is raw rate with chance-normalized rate (Eq.~\ref{eq:nleak}) in parenthesis ($\widetilde{\leak}$). 
\\The \textbf{bold} cell is each judge's \emph{most-following} placement, the prompt position at which it follows the wrong reference most often.}
\vspace{-1mm}
\label{tab:rq2-reference}
\begin{tabular}{@{}l cc cccc c cc cccc@{}}
\toprule
& \multicolumn{6}{c}{\textbf{Emotion} (RAVDESS, $K{=}8$, chance $0.14$)} & & \multicolumn{6}{c}{\textbf{Language} (FLEURS, $K{=}4$, chance $0.33$)} \\
\cmidrule(lr){2-7}\cmidrule(lr){9-14}
\textbf{Judge} & match & parse & before & after & system & rubric
& & match & parse & before & after & system & rubric \\
\midrule
Gemini-3-Flash
& 0.96 & 1.00 & 0.68\,(.63) & 0.63\,(.57) & \textbf{0.85\,(.82)} & 0.58\,(.51)
& & 1.00 & 1.00 & 0.00\,(.00) & 0.00\,(.00) & 0.00\,(.00) & 0.00\,(.00) \\
GPT-Audio
& 0.98 & 1.00 & 0.95\,(.94) & 0.45\,(.36) & 0.87\,(.85) & \textbf{0.99\,(.99)}
& & 1.00 & 0.97 & 0.12\,(.00) & 0.10\,(.00) & 0.12\,(.00) & \textbf{0.27\,(.00)} \\
Qwen3-Omni-Instruct
& 1.00 & 1.00 & 0.77\,(.73) & 0.87\,(.85) & 0.64\,(.58) & \textbf{0.97\,(.96)}
& & 1.00 & 1.00 & 0.00\,(.00) & 0.04\,(.00) & 0.00\,(.00) & \textbf{0.18\,(.00)} \\
Qwen3-Omni-Thinking
& 0.25 & 1.00 & 0.30\,(.18) & 0.49\,(.41) & 0.18\,(.04) & \textbf{0.57\,(.50)}
& & 1.00 & 1.00 & 0.15\,(.00) & 0.11\,(.00) & 0.09\,(.00) & \textbf{0.33\,(.00)} \\
Audio-Flamingo-3
& 0.83 & 1.00 & 0.16\,(.02) & \textbf{0.41\,(.31)} & 0.27\,(.15) & 0.33\,(.21)
& & 1.00 & 0.99 & 0.07\,(.00) & 0.60\,(.39) & 0.34\,(.01) & \textbf{0.89\,(.83)} \\
Voxtral-Small-24B
& 0.32 & 1.00 & 0.05\,(.00) & \textbf{0.25\,(.12)} & 0.04\,(.00) & 0.09\,(.00)
& & 0.20 & 0.10 & 0.08\,(.00) & 0.08\,(.00) & 0.04\,(.00) & 0.11\,(.00) \\
\bottomrule
\end{tabular}
\flushleft
\vspace{-4.5mm}
\end{table*}

% ==========================================================================
% SECTION 4 — RESULTS
% ==========================================================================
\section{Results}
\label{sec:results}
\vspace{-1mm}

We apply each protocol to two attributes and read each shortcut by whether it recurs when the attribute changes.
If a shortcut arises on both attributes, we treat it as a property of the protocol.
If it arises only when the judge cannot infer the attribute from audio, we treat it as capability-dependent.

\vspace{-1mm}

\subsection{Sufficiency: pointwise accuracy is not enough}
\label{sec:results-sanity}
Before the protocol audits, we confirm that aggregate pointwise accuracy is not by itself a sufficient diagnostic on our panel.
Prior work shows that LALMs can retain much of their full-audio accuracy when the audio is removed or contradicts the text~\cite{foo2026glitters,chen2026listen,billa2026modality,pang2026voxparadox}.

On the emotion classification task using RAVDESS, audio-only accuracy spans a wide range under the eight-way label set, from near the $0.125$ base rate of a transcript-only classifier up to $0.68$ for the strongest judge.
This spread reappears as a covariate in the tables shown in following subsections: judges that already engage the audio tend to resist the protocol shortcuts, and judges near the text baseline tend to follow them.
Thus, we treat audio-only accuracy as a background diagnostic and focus on the additional effect by each protocol.

\vspace{-1mm}

\subsection{RQ1: specialist copying depends on audio capability}
\label{sec:results-rq1}

We examine whether a structured blueprint of acoustic features supports acoustic reasoning.
We also test whether it becomes a channel for copying the embedded specialist block, and compare an attribute the judge can infer from audio with one it cannot.

Table~\ref{tab:rq1-blueprint-full} shows emotion (left) and language (right) results across the matched variants.
The two no-signal variants show the same pattern on both attributes: the \emph{skeleton} leaves every judge at the base rate, and inserting the \emph{true} block raises every judge to the specialist's own accuracy, whether or not the judge could have read the attribute unaided.

The variants diverge in the wrong-label columns: on emotion, \emph{wrong} reduces accuracy to near $0$ for five of six judges, since a judge that copies the wrong label scores zero against the truth. The most diagnostic condition, \emph{wrong-label$+$audio}, remains at or below $0.10$ for those five, with only Audio-Flamingo-3 partially recovering to $0.38$. On language, the same \emph{wrong-label$+$audio} column instead is near $1.00$ for four judges: audio evidence overrides the incorrect block, the opposite of the emotion result under an identical protocol.

\textbf{Language exceptions and capability dependence.} The two language exceptions make this dependence explicit.
Voxtral-Small cannot infer language from audio (audio-only $0.03$) as its \emph{wrong-label$+$audio} accuracy is furthest below ceiling ($0.72$). Qwen3-Omni-Thinking has partial language inference (audio-only $0.36$) as its \emph{wrong-label$+$audio} accuracy is at an intermediate $0.64$.
The four judges with near-ceiling language audio-only accuracy ($0.99$–$1.00$) are the four that reject the wrong block.
Capability and copying correspond: when the target attribute is recoverable from the audio, the wrong specialist block is ignored; when it is not, the block is followed. Therefore, the blueprint protocol does not introduce a fixed shortcut but introduces one only for attributes that a given judge cannot infer from audio. This explains why the emotion and language panels diverge under the same construction.

\textbf{Robustness across decoding seeds and best-of-$N$.} A $T{=}0.7$ seed sweep over the judges shows that Table~\ref{tab:rq1-blueprint-full} is stable: the maximum per-cell range is $0.038$, and 5-run majority vote shifts any cell by at most $0.008$.

% On \emph{wrong-label$+$audio}, GPT-Audio, Qwen3-Omni-Instruct, and Voxtral-Small output the identical wrong label on every call, so the $0.000$ entries are zero-entropy specialist copy rather than single-call noise.

\vspace{-1mm}

\subsection{RQ2: the placement-anchor recurs across attributes but is localized per judge and per placement}
\label{sec:results-rq2}

We examine whether a verbal reference calibrates the verdict or becomes an anchor whose strength depends on where in the prompt it appears, and whether that anchor behaves the same on an attribute the judge infers easily as one it does not.

Table~\ref{tab:rq2-reference} reports the reference-follow rate at four placements with the chance-normalized value in parentheses. 
On emotion the anchor is broad: five of six judges reach a chance-normalized rate above $0.30$ in at least one placement. Placement matters as much as the label, as the gap between a judge's most- and least-following slot reaches $0.54$ raw for GPT-Audio, and around $0.30$ for the Qwen3-Omni pair and Gemini-3-Flash.
The most-following slot is the rubric hint for GPT-Audio and the Qwen3-Omni pair and the system instruction for Gemini-3-Flash, so the channel is amplified by authoritative framing rather than by a fixed position.
Two judges resist on emotion: Audio-Flamingo-3 keeps every chance-normalized rate at or below $0.31$, and Voxtral-Small keeps three of four below chance.

On language, Audio-Flamingo-3 is the only judge whose reference-follow rate exceeds chance, reaching $0.83$ at the rubric slot---the same slot and the same judge-amplifying-framing pattern as on emotion, and the largest single above-chance rate in the table.
Every other judge remains at chance at all four placements. 
% At the rubric slot, GPT-Audio, Qwen3-Omni-Instruct, and Qwen3-Omni-Thinking all decline to a normalized $0.00$, so they would be interpreted as nonzero only without the chance correction.

The anchor thus recurs across attributes as a mechanism. The rubric slot triggers the largest reference-follow rate on both attributes, and Audio-Flamingo-3 anchors on both. On language, however, only Audio-Flamingo-3 reaches above-chance rate. The protocol opens the channel, while the magnitude depends on the judge and prompt placement.

\begin{table}[t]
\centering
\fontsize{8}{9}\selectfont
\setlength{\tabcolsep}{2.5pt}
\caption{RQ3 pairwise judging under AB/BA order swaps.\\
Each task reports \emph{lock} and \emph{g}: \emph{lock} is the same-slot rate, and \emph{g} is the gold-aligned rate over all trials in both orders. \\$p_A$ is the slot-A rate on the Identical control.}
\label{tab:rq3-pairwise}
\begin{tabular}{@{}l c cc cc cc cc@{}}
\toprule
& & \multicolumn{2}{c}{\textbf{Cleanness}} & \multicolumn{2}{c}{\textbf{Emotion}} & \multicolumn{2}{c}{\textbf{BVCC}} & \multicolumn{2}{c}{\textbf{Speaker}} \\
\cmidrule(lr){3-4}\cmidrule(lr){5-6}\cmidrule(lr){7-8}\cmidrule(lr){9-10}
\textbf{Judge} & $p_A$ & lock & $g$ & lock & $g$ & lock & $g$ & lock & $g$ \\
\midrule
\multicolumn{10}{@{}l}{\textsc{concat}} \\
Gemini-3-Flash       & 0.35 & 0.05 & \textbf{0.98} & 0.17 & \textbf{0.92} & 0.42 & \textbf{0.69} & 0.17 & \textbf{0.88} \\
GPT-Audio            & 0.10 & 0.97 & 0.51 & 0.36 & 0.54 & 0.76 & 0.52 & 0.82 & 0.51 \\
Qwen3-Omni-Instruct  & 0.88 & 0.38 & 0.26 & 0.27 & 0.87 & 0.55 & 0.58 & 0.49 & 0.64 \\
Qwen3-Omni-Thinking  & 1.00 & 1.00 & 0.50 & 1.00 & 0.50 & 1.00 & 0.50 & 1.00 & 0.50 \\
Audio-Flamingo-3     & 1.00 & 1.00 & 0.50 & 0.10 & 0.95 & 0.87 & 0.53 & 1.00 & 0.50 \\
Voxtral-Small-24B    & 0.83 & 0.60 & 0.60 & 0.47 & 0.60 & 0.87 & 0.48 & 0.80 & 0.51 \\
\midrule
\multicolumn{10}{@{}l}{\textsc{native}} \\
Gemini-3-Flash       & 0.48 & 0.20 & 0.85 & 0.21 & 0.90 & 0.40 & 0.67 & 1.00 & 0.50 \\
GPT-Audio            & 0.33 & 0.61 & 0.39 & 0.37 & 0.34 & 0.42 & 0.49 & 1.00 & 0.50 \\
\bottomrule
\end{tabular}
\vspace{-4mm}
\end{table}

\vspace{-1mm}

\subsection{RQ3: position-lock is format-sensitive, while engagement is task-dependent}

\label{sec:results-rq3}

We test whether pairwise A/B judging tracks audio identity or instead tracks the A/B slot and the input format, and whether the same pattern holds for a different comparative attribute, speaker similarity.

Table~\ref{tab:rq3-pairwise} reports a same-slot rate (\emph{lock}) and a gold-aligned rate (\emph{g}) for each judge.
Reading the two together partitions the panel into four modes.
\begin{itemize}[leftmargin=*,itemsep=2pt,topsep=2pt]
  \item \textbf{Engaged} (Gemini-3-Flash): lock remains low and $g$ remains high, so the evaluation setup can elicit audio-grounded verdicts from it.
  \item \textbf{Intermediate} (Qwen3-Omni-Instruct): it shows moderate lock, but its gold-aligned rate varies strongly by task, from $0.26$ (below chance) on Cleanness to $0.87$ on Emotion.
  \item \textbf{Locked} (GPT-Audio across all tasks; Voxtral-Small on BVCC and Speaker): lock remains at or above $0.76$ and $g$ falls to about $0.5$, the chance value a fixed slot produces. Voxtral-Small decreases to lock $0.47$--$0.60$ on Cleanness and Emotion, so the mode is task-dependent for this judge.
  \item \textbf{Fully locked} (Audio-Flamingo-3, Qwen3-Omni-Thinking): lock is $1.00$ and $g$ is $0.50$, since every verdict is assigned the same slot and matches the gold clip only when the gold clip occupies that slot.
\end{itemize}

Similar mode tendencies recur on the speaker triplets and the Cleanness control.
Gemini-3-Flash engages on both and Audio-Flamingo-3 and Qwen3-Omni-Thinking are fully locked on both, while GPT-Audio stays locked on both; Voxtral-Small and Qwen3-Omni-Instruct shift between the two tasks, so the lock structure recurs while engagement does not, even though the two tasks share no clips and ask different questions.
The $p_A$ column shows the locked rows are fixed slot preferences rather than prompt-template effects. 

% That the locked judges split between A-biased and B-biased, rather than sharing one slot, indicates the lock is a per-judge fixed effect that survives the order swap.

Two further effects appear within this split: 

\textbf{Engagement is task-specific.} On the Emotion control, Audio-Flamingo-3 reaches $g\!=\!0.95$ despite locking completely on Cleanness, and Qwen3-Omni-Instruct reaches $g\!=\!0.87$ despite partial locking on Cleanness. Thus, a single control would mis-describe a judge on another task.

\textbf{The format matters for one judge.} Switching to \textsc{native} multi-audio input lowers GPT-Audio's lock on Cleanness ($0.97\!\to\!0.61$) and BVCC ($0.76\!\to\!0.42$). Yet, $g$ remains near $0.5$, so the \textsc{concat} lock was in part a format artifact. Thus, removal does not by itself yield gold-aligned verdicts.

\begin{table}[t]
\centering
\fontsize{8}{9}\selectfont
\setlength{\tabcolsep}{3pt}
\caption{Cue-conflict grid on RAVDESS.
Cells are \emph{follow-cue}, the rate of emitting the wrong label; \textsc{none} is the audio-only baseline. \\The most-followed single channel per judge is in bold.}
\label{tab:cue}
\begin{tabular}{@{}l rrrr@{}}
\toprule
\textbf{Judge} & none & conf-spec & conf-ref & conf-both \\
\midrule
Gemini-3-Flash       & $0.44$ & $\mathbf{0.86}$ & $0.68$ & $0.87$ \\
GPT-Audio            & $0.23$ & $\mathbf{1.00}$ & $0.95$ & $1.00$ \\
Qwen3-Omni-Instruct  & $0.36$ & $\mathbf{1.00}$ & $0.77$ & $1.00$ \\
Qwen3-Omni-Thinking  & $0.13$ & $\mathbf{0.91}$ & $0.30$ & $0.95$ \\
Audio-Flamingo-3     & $0.70$ & $\mathbf{0.52}$ & $0.16$ & $0.68$ \\
Voxtral-Small-24B    & $0.32$ & $\mathbf{0.86}$ & $0.41$ & $0.88$ \\
\bottomrule
\end{tabular}
\vspace{-4mm}
\flushleft
\end{table}

\subsection{Synthesis: the specialist block is the most influential}
\label{sec:results-synthesis}

As blueprint and reference protocols each introduce a shortcut in isolation, we examine which exerts greater influence when both compete on the same audio.

Table~\ref{tab:cue} reports the grid.
The specialist block (\textsc{conf-spec}) is followed more often than the verbal reference (\textsc{conf-ref}) in every row.
GPT-Audio and Qwen3-Omni-Instruct reach the ceiling under \textsc{conf-spec}.
The difference is substantial for Gemini-3-Flash ($0.86$ vs.\ $0.68$).
Wider differences are observed for Audio-Flamingo-3 ($0.52$ vs.\ $0.16$) and Qwen3-Omni-Thinking ($0.91$ vs.\ $0.30$), the two judges with the lowest \textsc{conf-ref} in the grid and thus the most resistant to the reference channel.
For them, the reference is followed less than half as often as the specialist block on the same clip.

The matched \emph{wrong$+$audio} and wrong-reference columns of Tables~\ref{tab:rq1-blueprint-full} and~\ref{tab:rq2-reference} show the same ordering, so it holds across two independent sweeps. Supplying both channels at once (\textsc{conf-both}) adds little over the specialist block alone.
The two ceiling judges cannot increase further: Gemini-3-Flash, Audio-Flamingo-3, and Qwen3-Omni-Thinking rise only to $0.87$, $0.68$, and $0.95$.
Once the specialist block is present, the reference contributes minimally toward the same wrong label.

\section{Conclusion} 
\label{sec:conclusion} 

We audited three LALM-as-a-judge deployment protocols for speech evaluation: feature-blueprint judging, reference-conditioned judging, and pairwise A/B comparison. Across six closed and open LALMs, the results show that judge validity depends on the model--protocol pair, not on the model alone. Reference-conditioned judging becomes a prompt-placement-dependent anchor, and pairwise judging often becomes a slot- or format-dependent comparison. Because both patterns recur on a second attribute, we treat them as protocol-level effects. 

Feature-blueprint judging shows a different pattern. Judges copy a wrong specialist label on emotion even when audio is present, but reject the label on language when they can infer the attribute from audio. This contrast suggests a capability-dependent shortcut: the blueprint channel becomes unreliable when the judge cannot verify the specialist field acoustically. 

These results indicate that aggregate agreement with human labels is insufficient for diagnosing protocol-induced behavior. Each deployment protocol should be paired with its own shortcut probe: wrong-label specialist blocks for blueprint judges, placement sweeps for reference judges, and order-swapped positive controls with concat/native format contrasts for pairwise judges. 
Future work should extend the audit to a wider panel of judges and to harder attributes, to test how far the protocol-level and capability-dependent split generalizes. 
% Beyond diagnostics, the matched probes could also serve as training signals that fine-tune judges to stay audio-grounded under each protocol.

\bibliographystyle{IEEEtran}
\bibliography{bib/tts}

@inproceedings{goel2025audioflamingo3,
  title     = {Audio Flamingo 3: Advancing Audio Intelligence with Fully Open
               Large Audio Language Models},
  author    = {Goel, Arushi and Ghosh, Sreyan and Kim, Jaehyeon and Kumar, Sonal and
               Kong, Zhifeng and Lee, Sang-gil and Yang, Chao-Han Huck and
               Duraiswami, Ramani and Manocha, Dinesh and Valle, Rafael and
               Catanzaro, Bryan},
  booktitle = {Proceedings of Advances in Neural Information Processing Systems (NeurIPS)},
  year      = {2025},
  note      = {Spotlight},
}

@inproceedings{wu2024claira,
  title     = {{CLAIR-A}: Leveraging Large Language Models to Judge Audio Captions},
  author    = {Wu, Tsung-Han and Gonzalez, Joseph E. and Darrell, Trevor and
               Chan, David M.},
  booktitle = {Proceedings of the IEEE Automatic Speech Recognition and
               Understanding Workshop (ASRU)},
  year      = {2025},
}

@inproceedings{wang2025speechjudge,
  title     = {{SpeechLLM}-as-Judges: Towards General and Interpretable Speech
               Quality Evaluation},
  author    = {Wang, Hui and others},
  booktitle = {Proceedings of the Annual Meeting of the Association for
               Computational Linguistics (ACL)},
  year      = {2026},
}

@inproceedings{chen2026listen,
  title     = {Do Audio {LLM}s Really {LISTEN}, or Just Transcribe? Measuring
               Lexical vs.\ Acoustic Emotion Cues Reliance},
  author    = {Chen, Jingyi and Guo, Zhimeng and Chun, Jiyun and Wang, Pichao and
               Perrault, Andrew and Elsner, Micha},
  booktitle = {Proceedings of the Conference of the European Chapter of the
               Association for Computational Linguistics (EACL)},
  year      = {2026},
}

@inproceedings{ivry2026lalmjudge,
  title     = {{LALM-as-a-Judge}: Benchmarking Large Audio-Language Models for
               Safety Evaluation in Multi-Turn Spoken Dialogues},
  author    = {Ivry, Amir and Watanabe, Shinji},
  booktitle = {Proceedings of the International Conference on Machine
               Learning (ICML)},
  year      = {2026},
}

@inproceedings{zheng2023judging,
  title     = {Judging {LLM}-as-a-Judge with {MT-Bench} and Chatbot Arena},
  author    = {Zheng, Lianmin and Chiang, Wei-Lin and Sheng, Ying and
               Zhuang, Siyuan and Wu, Zhanghao and Zhuang, Yonghao and
               Lin, Zi and Li, Zhuohan and Li, Dacheng and Xing, Eric P. and
               Zhang, Hao and Gonzalez, Joseph E. and Stoica, Ion},
  booktitle = {Proceedings of Advances in Neural Information Processing
               Systems (NeurIPS)},
  volume    = {36},
  pages     = {46595--46623},
  year      = {2023},
}

@inproceedings{chan2023clair,
  title     = {{CLAIR}: Evaluating Image Captions with Large Language Models},
  author    = {Chan, David M. and Petryk, Suzanne and Gonzalez, Joseph E. and
               Darrell, Trevor and Canny, John},
  booktitle = {Proceedings of the 2023 Conference on Empirical Methods in
               Natural Language Processing (EMNLP)},
  pages     = {13638--13646},
  year      = {2023},
}

@inproceedings{chen2025alld,
  title     = {Audio Large Language Models Can Be Descriptive Speech Quality
               Evaluators},
  author    = {Chen, Chen and Hu, Yuchen and Wang, Siyin and Wang, Helin and
               Chen, Zhuo and Zhang, Chao and Yang, Chao-Han Huck and
               Chng, Eng Siong},
  booktitle = {Proceedings of the International Conference on Learning
               Representations (ICLR)},
  year      = {2025},
}

@inproceedings{manku2025emergent,
  title     = {{EmergentTTS-Eval}: Evaluating {TTS} Models on Complex Prosodic,
               Expressiveness, and Linguistic Challenges Using Model-as-a-Judge},
  author    = {Manku, Ruskin Raj and Tang, Yuzhi and Shi, Xingjian and
               Li, Mu and Smola, Alex},
  booktitle = {Proceedings of Advances in Neural Information Processing
               Systems (NeurIPS)},
  year      = {2025},
}

@inproceedings{chandra2026trace,
  title     = {Hearing Between the Lines: Unlocking the Reasoning Power of
               {LLMs} for Speech Evaluation},
  author    = {Chandra, Arjun and Miller, Kevin and Ravichandran, Venkatesh and
               Papayiannis, Constantinos and Saligrama, Venkatesh},
  booktitle = {Proceedings of the Findings of the Association for Computational
               Linguistics: EACL 2026},
  year      = {2026},
}

@inproceedings{ma2023emotion2vec,
  title     = {emotion2vec: Self-Supervised Pre-Training for Speech Emotion
               Representation},
  author    = {Ma, Ziyang and Zheng, Zhisheng and Ye, Jiaxin and Li, Jinchao and
               Gao, Zhifu and Zhang, Shiliang and Chen, Xie},
  booktitle = {Proceedings of the Findings of the Association for Computational
               Linguistics: ACL 2024},
  pages     = {15747--15760},
  year      = {2024},
}

@inproceedings{cooper2023bvcc,
  title     = {The {VoiceMOS} {C}hallenge 2022},
  author    = {Cooper, Erica and Huang, Wen-Chin and Tsao, Yu and Wang, Hsin-Min and
               Toda, Tomoki and Yamagishi, Junichi},
  booktitle = {Proceedings of INTERSPEECH},
  year      = {2022},
}

@inproceedings{pang2026voxparadox,
  title     = {Do Audio {LLM}s Listen or Read? Analyzing and Mitigating
               Paralinguistic Failures with {VoxParadox}},
  author    = {Pang, Jiacheng and Chaubey, Ashutosh and Soleymani, Mohammad},
  booktitle = {Proceedings of the International Conference on Machine
               Learning (ICML)},
  year      = {2026},
}

@inproceedings{wang2024fairevaluators,
  title     = {Large Language Models are not Fair Evaluators},
  author    = {Wang, Peiyi and Li, Lei and Chen, Liang and Cai, Zefan and Zhu, Dawei and
               Lin, Binghuai and Cao, Yunbo and Kong, Lingpeng and Liu, Qi and
               Liu, Tianyu and Sui, Zhifang},
  booktitle = {Proceedings of the 62nd Annual Meeting of the Association for
               Computational Linguistics (Volume 1: Long Papers)},
  pages     = {9440--9450},
  year      = {2024},
  address   = {Bangkok, Thailand},
  publisher = {Association for Computational Linguistics},
  doi       = {10.18653/v1/2024.acl-long.511},
}

@inproceedings{ye2025justice,
  title     = {Justice or Prejudice? Quantifying Biases in {LLM}-as-a-Judge},
  author    = {Ye, Jiayi and Wang, Yanbo and Huang, Yue and Chen, Dongping and
               Zhang, Qihui and Moniz, Nuno and Gao, Tian and Geyer, Werner and
               Huang, Chao and Chen, Pin-Yu and Chawla, Nitesh V. and Zhang, Xiangliang},
  booktitle = {Proceedings of the International Conference on Learning
               Representations (ICLR)},
  year      = {2025},
}

@inproceedings{fanous2025syceval,
  title     = {{SycEval}: Evaluating {LLM} Sycophancy},
  author    = {Fanous, Aaron and others},
  booktitle = {Proceedings of the AAAI/ACM Conference on AI, Ethics, and
               Society (AIES)},
  pages     = {893--900},
  year      = {2025},
  doi       = {10.1609/aies.v8i1.36598},
}

@inproceedings{sharma2024sycophancy,
  title     = {Towards Understanding Sycophancy in Language Models},
  author    = {Sharma, Mrinank and Tong, Meg and Korbak, Tomasz and Duvenaud, David and
               Askell, Amanda and Bowman, Samuel R. and Durmus, Esin and
               Hatfield-Dodds, Zac and Johnston, Scott R. and Kravec, Shauna M. and others},
  booktitle = {Proceedings of the International Conference on Learning
               Representations (ICLR)},
  year      = {2024},
}

@article{huang2025instructtts,
  title   = {{InstructTTSEval}: Benchmarking Complex Natural-Language
             Instruction Following in Text-to-Speech Systems},
  author  = {Huang, Kexin and Tu, Qian and Fan, Liwei and Yang, Chenchen and
             Zhang, Dong and Li, Shimin and Fei, Zhaoye and Cheng, Qinyuan and
             Qiu, Xipeng},
  journal = {arXiv preprint arXiv:2506.16381},
  year    = {2025},
}

@article{livingstone2018ravdess,
  title     = {The {R}yerson {A}udio-{V}isual {D}atabase of {E}motional
               {S}peech and {S}ong ({RAVDESS}): A dynamic, multimodal
               set of facial and vocal expressions in {N}orth {A}merican
               {E}nglish},
  author    = {Livingstone, Steven R. and Russo, Frank A.},
  journal   = {PLoS ONE},
  volume    = {13},
  number    = {5},
  pages     = {e0196391},
  year      = {2018},
  doi       = {10.1371/journal.pone.0196391},
}

@article{chu2024qwen2audio,
  title={Qwen2-Audio Technical Report},
  author={Chu, Yunfei and Xu, Jin and Yang, Qian and Wei, Haojie and Wei, Xipin and Guo, Zhifang and Leng, Yichong and Lv, Yuanjun and He, Jinzheng and Lin, Junyang and Zhou, Chang and Zhou, Jingren},
  journal={arXiv preprint arXiv:2407.10759},
  year={2024}
}

@article{xu2025qwen25omni,
  title={Qwen2.5-Omni Technical Report},
  author={Xu, Jin and Guo, Zhifang and He, Jinzheng and Hu, Hangrui and He, Ting and Bai, Shuai and Chen, Keqin and Wang, Jialin and Yang, Fan and Dang, Kai and others},
  journal={arXiv preprint arXiv:2503.20215},
  year={2025}
}

@article{xu2025qwen3omni,
  title={Qwen3-Omni Technical Report},
  author={Xu, Jin and Guo, Zhifang and Hu, Hangrui and Chu, Yunfei and Wang, Xiong and He, Jinzheng and Wang, Yuxuan and Shi, Xian and He, Ting and Zhu, Xinfa and others},
  journal={arXiv preprint arXiv:2509.17765},
  year={2025}
}

@article{liu2025voxtral,
  title={Voxtral},
  author={Liu, Alexander H. and others},
  journal={arXiv preprint arXiv:2507.13264},
  year={2025}
}

@article{wang2025audiojudge,
  title   = {{AudioJudge}: Understanding What Works in Large Audio Model Based
             Speech Evaluation},
  author  = {Manakul, Potsawee and Gan, Woody Haosheng and Ryan, Michael J. and
             Khan, Ali Sartaz and Sirichotedumrong, Warit and
             Pipatanakul, Kunat and Held, William and Yang, Diyi},
  journal = {arXiv preprint arXiv:2507.12705},
  year    = {2025},
}

@article{foo2026glitters,
  title   = {All That Glitters Is Not Audio: Rethinking Text Priors and Audio
             Reliance in Audio-Language Evaluation},
  author  = {Foo, Leonardo Haw-Yang and Yang, Chih-Kai and Li, Chen-An and
             Lu, Ke-Han and Lee, Hung-yi},
  journal = {arXiv preprint arXiv:2604.24401},
  year    = {2026},
}

@article{billa2026modality,
  title   = {When Audio-{LLM}s Don't Listen: A Cross-Linguistic Study of
             Modality Arbitration},
  author  = {Billa, Jayadev},
  journal = {arXiv preprint arXiv:2602.11488},
  year    = {2026},
}

@article{eyben2016gemaps,
  title   = {The {Geneva} Minimalistic Acoustic Parameter Set ({GeMAPS}) for Voice Research and Affective Computing},
  author  = {Eyben, Florian and Scherer, Klaus R. and Schuller, Bj{\"o}rn W. and Sundberg, Johan and Andr{\'e}, Elisabeth and Busso, Carlos and Devillers, Laurence Y. and Epps, Julien and Laukka, Petri and Narayanan, Shrikanth S. and Truong, Khiet P.},
  journal = {IEEE Transactions on Affective Computing},
  volume  = {7},
  number  = {2},
  pages   = {190--202},
  year    = {2016},
  doi     = {10.1109/TAFFC.2015.2457417},
}

@article{geirhos2020shortcut,
  title   = {Shortcut Learning in Deep Neural Networks},
  author  = {Geirhos, Robert and Jacobsen, J{\"o}rn-Henrik and Michaelis, Claudio and Zemel, Richard and Brendel, Wieland and Bethge, Matthias and Wichmann, Felix A.},
  journal = {Nature Machine Intelligence},
  volume  = {2},
  number  = {11},
  pages   = {665--673},
  year    = {2020},
  doi     = {10.1038/s42256-020-00257-z},
}

@inproceedings{shah2020pitfalls,
  title     = {The Pitfalls of Simplicity Bias in Neural Networks},
  author    = {Shah, Harshay and Tamuly, Kaustav and Raghunathan, Aditi and Jain, Prateek and Netrapalli, Praneeth},
  booktitle = {Proceedings of Advances in Neural Information Processing Systems (NeurIPS)},
  volume    = {33},
  pages     = {9573--9585},
  year      = {2020},
}

@inproceedings{tang2023lazy,
  title     = {Large Language Models Can be Lazy Learners: Analyze Shortcuts in In-Context Learning},
  author    = {Tang, Ruixiang and Kong, Dehan and Huang, Longtao and Xue, Hui},
  booktitle = {Findings of the Association for Computational Linguistics: ACL 2023},
  pages     = {4645--4657},
  year      = {2023},
  address   = {Toronto, Canada},
  publisher = {Association for Computational Linguistics},
  doi       = {10.18653/v1/2023.findings-acl.284},
}

@inproceedings{nagrani2017voxceleb,
  title     = {{VoxCeleb}: A Large-Scale Speaker Identification Dataset},
  author    = {Nagrani, Arsha and Chung, Joon Son and Zisserman, Andrew},
  booktitle = {Proceedings of INTERSPEECH},
  pages     = {2616--2620},
  year      = {2017},
}

@inproceedings{radford2023whisper,
  title     = {Robust Speech Recognition via Large-Scale Weak Supervision},
  author    = {Radford, Alec and Kim, Jong Wook and Xu, Tao and Brockman, Greg and McLeavey, Christine and Sutskever, Ilya},
  booktitle = {Proceedings of the International Conference on Machine Learning (ICML)},
  series    = {Proceedings of Machine Learning Research},
  volume    = {202},
  pages     = {28492--28518},
  year      = {2023},
  publisher = {PMLR},
}

@inproceedings{desplanques2020ecapa,
  title     = {{ECAPA-TDNN}: Emphasized Channel Attention, Propagation and Aggregation in {TDNN} Based Speaker Verification},
  author    = {Desplanques, Brecht and Thienpondt, Jenthe and Demuynck, Kris},
  booktitle = {Proc. Interspeech},
  pages     = {3830--3834},
  year      = {2020},
  doi       = {10.21437/Interspeech.2020-2650},
}

\end{document}